\documentclass[amsfonts,aps,nofootinbib,preprint,superscriptaddress]{revtex4}

\usepackage{graphicx}

\begin{document}

\title{Superspace evaluation of the two-loop effective potential for the O'Raifeartaigh model}

\author{M. C. B. Abdalla{\footnote{mabdalla@ift.unesp.br}}}
\affiliation{Instituto de F\'{\i}sica Te\'{o}rica, UNESP - Universidade Estadual Paulista, Rua Dr. Bento Teobaldo Ferraz 271, Bloco II, Barra-Funda, Caixa Postal 70532-2, 01156-970, S\~{a}o Paulo, SP, Brazil }
\author{J. A. Helay\"{e}l-Neto{\footnote{helayel@cbpf.br}}}
\affiliation{Centro Brasileiro de Pesquisas F\'isicas, Rua Dr. Xavier Sigaud 150, 
Urca, Rio de Janeiro, RJ, 22290-180, Brazil}
\author{Daniel L. Nedel{\footnote{daniel.nedel@unipampa.edu.br}}}
\affiliation{Universidade Federal do Pampa, 
Rua Carlos Barbosa S/N, Bairro Get\'ulio Vargas, 96412-420, Bag\'e, RS, Brazil}
\author{Carlos R. Senise Jr.{\footnote{senise@ift.unesp.br}}}
\affiliation{Instituto de F\'{\i}sica Te\'{o}rica, UNESP - Universidade Estadual Paulista, Rua Dr. Bento Teobaldo Ferraz 271, Bloco II, Barra-Funda, Caixa Postal 70532-2, 01156-970, S\~{a}o Paulo, SP, Brazil }

\begin{abstract}
All-order spurion-corrected superpropagators and superfield Feynman rules are employed to systematically compute a two-loop corrected effective potential for the O'Raifeartaigh model, that realizes spontaneous supersymmetry breaking. Though the shifted superpropagators are rather nontrivial, superspace techniques may be suitably extended and confirm their efficacy in computing radiative corrections even when supersymmetry breakdown occurs. 
\end{abstract}

\maketitle

The effective potential plays a very important role in determining the nature of the vacuum in
quantum  field theories. It  allows the calculation of the vacuum expectation values in the true vacuum state
of a theory with spontaneous symmetry breaking.

In the case of supersymmetric theories, the effective potential can be directly calculated in superspace, by using supergraphs. In general, the supereffective action is described by two functions of the chiral and the antichiral superfields; one is required to be a holomorphic function and the other one, called K\"{a}hler potential, less constrained, is required to be just a real function. The holomorphic part of the  superpotential is very constrained, which is reflected in various non–renormalization theorems, leading to results to all orders in perturbation theory \cite{nonrenormalization}, and even nonperturbative results in some cases \cite{SWn=2}. For models with spontaneous supersymmetry (SUSY) breaking, the effective potential can be calculated by using superspace techniques even if soft explicit breaking terms are introduced in the superspace action; this yields spurion insertions. These terms have been carefully classified and studied by Girardello and Grisaru \cite{Girardello-Grisaru}. This approach to study SUSY breaking is very powerful because it leaves most of the supersymmetric structure intact. In fact, as the full supersymmetric and the supersymmetry breaking terms are represented as interactions in superspace, the renormalization can be performed systematically directly in superspace.
 
In general, the background field method is adopted to calculate the effective potential. In this method,  the scalar fields of the theory are each separated into a constant classical background plus quantum fluctuations. 
Using this approach, the effective potential is equal to the tree-level potential in the classical background,
plus the sum of one-particle-irreducible connected vacuum graphs with field-dependent masses and couplings.  In superspace the  vacuum supergraphs are identically zero, owing to the Berezin integrals. However, when soft breaking terms are present, the propagators have a nontrivial $\theta$-dependent part and the vacuum supergraphs do contribute \cite{Helayel}. 

Using superspace spurion techniques, the superpotential for the simplest O'Raifeartaigh model for spontaneous breaking of supersymmetry was calculated in \cite{Helayel} at the one-loop level. Although the O'Raifeartaigh model for SUSY breaking has been fairly well studied, this kind of model has recently received renewed attention in view of the \textit{R} symmetry, which plays an important role in SUSY breaking \cite{NS}. 
Though the simplest original O'Raifeartaigh model \cite{O'Raifeartaigh} does not spontaneously break \textit{R} symmetry, generalized O'Raifeartaigh models, which spontaneously violate \textit{R} symmetry, have been built up \cite{IS,ISS1,ISS2}. In these generalized O'Raifeartaigh models, \textit{R} symmetry is spontaneously broken by the pseudomoduli, which are charged (under \textit{R} symmetry) and acquire a nonzero vacuum expectation value via radiative corrections incorporated into the effective potential \cite{Shih,Marques} at one loop. So, it is very important to develop methods that account for higher-loop corrections to the effective potential of O'Raifeartaigh-type models. In \cite{twoloop}, a component-field approach was used to study \textit{R} symmetry breaking at two loops.

The main goal of this paper is to show that the superspace approach, with spurion insertions,  can be a powerful tool to derive higher-loop corrections to the effective potential of O'Raifeartaigh-type models. To this end, we used the spurion techniques developed in \cite{Helayel}, \cite{Nibbelink} to calculate the effective potential for the simplest O'Raifeartaigh model at two loops directly in superspace. As we are going to use a superspace field method, we have just one kind of two-loop vacuum diagram to calculate. Though this approach drastically reduces the number of individual diagrams with respect to a component-field calculation, a drawback is that the propagators have a nontrivial dependence on the spurion insertions and so the vacuum diagram involves an infinite sum of the spurion terms. The spurion insertions can however be summed up to all orders. The basic techniques for dealing with such a problem were developed in \cite{Nibbelink}, when spurion operators were introduced, which satisfy a Clifford algebra and simplify the computations. Here, we further extend these techniques to calculate two-loop vacuum supergraphs. At this point, we would like to recall that the application of superfield methods to derive vacuum diagrams in superspace to one- and two-loop approximations in connection with the renormalization of the K\"{a}hler potential has already been considered in the literature \cite{Kpotential}.

The paper is organized according to the following outline: in Sec. I, we present the model, derive the Feynman rules and calculate the one-loop effective potential; in Sec. II, we calculate the two-loop vacuum diagrams and show that, after integrating over the superspace coordinates, the remaining expressions are written in terms of usual space-time integrals; in Sec. III, we solve the space-time integrals and obtain the final expression for the renormalized two-loop effective potential. Concluding remarks are finally cast in Sec. IV. There follow two Appendices. In Appendix A, we present useful algebraic relations in superspace and explicitly calculate two of the superspace integrals presented in Sec. II. We collect in the Appendix B some of the intermediate steps of our supergraph calculations, which may be helpful to get the final expressions reported in Sec. III. 


\section{The one-loop effective potential}
The Lagrangian for the O'Raifeartaigh model \cite{O'Raifeartaigh} is as follows:
\begin{equation}
{\mathcal L}=\int d^{4}\theta\bar{\Phi}_{i}\Phi_{i}-\left[\int d^{2}\theta\left(\xi\Phi_{0}+m\Phi_{1}\Phi_{2}+g\Phi_{0}\Phi_{1}^{2}\right)+h.c.\right] \ \ \ , \ \ \ i=0,1,2 \ . \label{O'Raifeartaigh}
\end{equation}

The chiral superfields, $\Phi_{i}$, are given by
\begin{equation}
\Phi_{i}(x,\theta,\bar{\theta})=\exp(i\theta\sigma^{\mu}\bar{\theta}\partial_{\mu})(z_{i}+\theta\psi_{i}+\theta^{2}h_{i}) \ , \label{superfield}
\end{equation}
where $z$ is the scalar, $\psi$ is the spin-1/2, and $h$ is the scalar auxiliary component fields.

In order to show the SUSY breaking in the model, and for later convenience, it is necessary to study the potential of the model described by (\ref{O'Raifeartaigh}). The scalar potential is given in terms of the auxiliary fields, $h_{i}$, by
\begin{equation}
V_{S}=h_{i}h^{\ast}_{i}=|\xi+gz_{1}^{2}|^{2}+|mz_{2}+2gz_{0}z_{1}|^{2}+|mz_{1}|^{2} \ . \label{V scalar}
\end{equation}
Writing the vacuum expectation values (vev) of the scalar fields as $\langle z_{i}\rangle$, the minimum of (\ref{V scalar}) in the tree approximation is \cite{Helayel}
\begin{equation}
\langle z_{0}\rangle=y \ \ \ , \ \ \ \langle z_{1}\rangle=x \ \ \ , \ \ \ \langle z_{2}\rangle=-\frac{2g}{m}xy \ , \label{z values}
\end{equation}
where $y$ is completely undetermined (flat direction) and $x$ is real and obeys the equation $x(m^{2}+2g\xi+2g^{2}x^{2})=0$. At this minimum,
\begin{equation}
V_{S}^{min}=(\xi+gx^{2})^{2}+(mx)^{2}=V(x) \ . \label{V scalar min}
\end{equation}
Solving the equations of motion for the auxiliary fields, we obtain
\begin{equation}
\langle h_{0}\rangle=\Delta \ \ \ , \ \ \ \langle h_{1}\rangle=0 \ \ \ , \ \ \ \langle h_{2}\rangle=mx \ , \label{h values} 
\end{equation} 
where $\Delta=\xi+gx^{2} \ .$ The relations above show that the O'Raifeartaigh model, described by (\ref{O'Raifeartaigh}), in fact breaks SUSY, since we have nonvanishing vev's for some of the auxiliary fields.

The values of $x$ and $y$, and consequently of the vev's $\langle z_{i}\rangle, \ \langle h_{i}\rangle$, are related to symmetric or nonsymmetric phases of the system. We do not intend to discuss these features in the present work, and for this we refer the reader to the references \cite{Helayel,Huq}. For our purposes, it is sufficient to know that $y=0$ due to one-loop corrections.

Since in the following we shall work around the ground state, we take $\langle h_{0}\rangle=\Delta$, $\langle z_{1}\rangle=x$ and $\langle h_{2}\rangle=mx$, and shift the superfields as a sum of a quantum superfield ($\phi_0$, $\phi_1$ and $\phi_2$) plus the related classical value:
\begin{eqnarray}
\Phi_{0}&=&\phi_{0}+\theta^{2}\Delta \ , \nonumber\\
\Phi_{1}&=&\phi_{1}+x \ , \nonumber\\
\Phi_{2}&=&\phi_{2}+\theta^{2}mx \ . \label{phi shifted}
\end{eqnarray}

Inserting the shifted fields into (\ref{O'Raifeartaigh}) and expanding the action around the superspace classical configuration up to third order in the quantum fields, we obtain
\begin{equation}
{\mathcal L}=\int d^{4}\theta\bar{\phi}_{i}\phi_{i}-\left[\int d^{2}\theta\left(\xi\phi_{0}+m\phi_{1}\phi_{2}+2gx\phi_{0}\phi_{1}+g\Delta\theta^{2}\phi_{1}^{2}+g\phi_{0}\phi_{1}^{2}\right)+h.c.\right] \ , \label{shifted O'Raifeartaigh}
\end{equation}

The appearance of terms proportional to $\theta$ and $\bar{\theta}$ in (\ref{shifted O'Raifeartaigh}) signals the explicit breaking of SUSY and naturally arise when spontaneous SUSY breaking is studied in superspace. These are the spurion interactions classified in \cite{Girardello-Grisaru}.

The propagators can be derived if we invert the wave operator in the quadratic part of the Lagrangian. By using the techniques developed in \cite{Helayel}, we find that they are given as follows:
\begin{eqnarray}
\langle\phi_{0}\bar{\phi_{0}}\rangle&=&(k^{2}+m^{2})A(k)\delta^{4}_{12}+(2gx)^{2}(2g\Delta)^{2}B(k)\theta_{1}^{2}\bar{\theta}_{1}^{2}\delta_{12}^{4} \ ; \nonumber\\
\langle\phi_{0}\bar{\phi_{1}}\rangle&=&(2gx)(2g\Delta) C(k)\frac{1}{16}D_{1}^{2}\bar{D}_{1}^{2}\theta_{1}^{2}\delta_{12}^{4} \ ; \nonumber\\
\langle\phi_{1}\bar{\phi_{1}}\rangle&=&E(k)\delta_{12}^{4}+(2g\Delta)^{2}B(k)\frac{1}{16}D_{1}^{2}\theta_{1}^{2}\bar{\theta}_{1}^{2}\bar{D}_{1}^{2}\delta_{12}^{4} \ ; \nonumber\\
\langle\phi_{0}\phi_{0}\rangle&=&-(2gx)^{2}(2g\Delta)C(k)\frac{1}{4}D_{1}^{2}\theta_{1}^{2}\delta_{12}^{4} \ ; \nonumber\\
\langle\phi_{0}\phi_{1}\rangle&=&(2gx)A(k)\frac{1}{4}D_{1}^{2}\delta_{12}^{4}-(2gx)(2g\Delta)^{2}B(k)\frac{1}{4}\theta_{1}^{2}\bar{\theta}_{1}^{2}D_{1}^{2}\delta_{12}^{4} \ ; \nonumber\\
\langle\phi_{1}\phi_{1}\rangle&=&(2g\Delta)F(k)\frac{1}{4}\bar{\theta}_{1}^{2}D_{1}^{2}\delta_{12}^{4} \ , \label{propagators}
\end{eqnarray}
where
\begin{eqnarray*}
A(k)&=&\frac{1}{k^{2}\left(k^{2}+m^{2}+4g^{2}x^{2}\right)} \ \ , \\ B(k)&=&\frac{1}{\left(k^{2}+m^{2}+4g^{2}x^{2}\right)\left[\left(k^{2}+m^{2}+4g^{2}x^{2}\right)^{2}-4g^{2}\Delta^{2}\right]} \ \ , \\
C(k)&=&\frac{1}{k^{2}\left[\left(k^{2}+m^{2}+4g^{2}x^{2}\right)^{2}-4g^{2}\Delta^{2}\right]} \ \ , \\ 
E(k)&=&\frac{1}{k^{2}+m^{2}+4g^{2}x^{2}} \ \ , \\ 
F(k)&=&\frac{1}{\left(k^{2}+m^{2}+4g^{2}x^{2}\right)^{2}-4g^{2}\Delta^{2}} \ \ ,
\end{eqnarray*}
and $\delta^{4}_{12}=\delta^{4}(\theta_1-\theta_2)=\delta^{4}_{21}$. We do not consider the propagators involving the $\phi_{2}$ superfield, since they do not appear in the vacuum diagrams.

We can also write the Feynman rule for the interaction term $\phi_{0}\phi_{1}^{2}$:
\begin{equation}
\phi_{0}\phi_{1}^{2} \ \hbox{vertex} \ :-2g\int d^{4}\theta \ . \label{vextex}
\end{equation}
 
The quantum contributions to the effective potential can now be evaluated  in superspace by calculating
the one-particle-irreducible connected vacuum supergraphs using the Feynman rules defined in (\ref{propagators}) and (\ref{vextex}). Note that the propagators have a nontrivial dependence on the spurion interactions. 

After expanding the supergenerating functional, the effective potential can be expressed as follows:
\begin{equation}
{\mathcal V}_{eff}={\mathcal V}_{eff}^{(0)}+ \frac{1}{(4\pi)^{2}}{\mathcal V}_{eff}^{(1)}+ \frac{1}{(4\pi)^4}{\mathcal V}_{eff}^{(2)}+... \ \ , \label{V expansion}
\end{equation} 
where ${\mathcal V}_{eff}^{(n)}$ represents the \textit{n}-loop correction. The one-loop vacuum diagram shown in Fig. 1 comes from the quadratic part of the action expansion, while the two-loop vacuum diagrams come from the three-quantum field terms of the action expansion.
\begin{eqnarray*}
\begin{picture}(20,0) \thicklines
\put(15,-5){\circle{40}}
\end{picture}
\end{eqnarray*}

\vspace{0,1cm}
\begin{center}
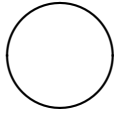
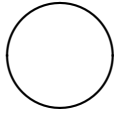
Fig. 1: One-loop vacuum diagram. 
\end{center}
By calculating the Gaussian integral in superspace and using the definition of the super effective action, the one-loop vacuum diagram is written as a supertrace and one can adopt the same techniques developed in \cite{Nibbelink} to deal with the spurion insertions, 
\begin{equation}
{\mathcal V}_{eff}^{(1)}=-\frac{1}{2}\int d^{4}\theta_{12}\delta^{4}_{21}Tr\ln[P^{T}K]\delta^{4}_{21} \ , \label{order 1}
\end{equation}
where $d^{4}\theta_{12}=d^{4}\theta_{1}d^{4}\theta_{2}$ and the notation $Tr$ refers to the trace over the chiral multiplets in the real basis defined by the vector $(\Phi^T,\bar{\Phi})^T$. $P$ is the matrix defined by the chiral projectors $P_{+}=\frac{\bar{D}^{2}D^{2}}{16\Box}$ and $P_{-}=\frac{D^{2}\bar{D}^{2}}{16\Box}$ as
\begin{equation}
P=\left(\begin{array}{cc}
          0 & P_{-} \\
          P_{+} & 0
          \end{array}\right) \ , \label{P}
\end{equation}
and 
\begin{equation}
K=\left(\begin{array}{cc}
          \left(AP_{-}+B\frac{1}{\Box^{1/2}}\eta_{-}\right)\frac{D^2}{4\Box} & \textbf{1}_{3\times3} \\
          \textbf{1}_{3\times3} & \left(\bar{A}P_{+}+\bar{B}\frac{1}{\Box^{1/2}}\bar{\eta}_{+}\right)\frac{\bar{D}^2}{4\Box}
          \end{array}\right) \ , \label{K}
\end{equation}
with
\begin{equation}
A=\left(\begin{array}{ccc}
          0 & 2gx & 0 \\
          2gx & 0 & m \\
          0 & m & 0
          \end{array}\right) \ \ , \ \ 
B=\left(\begin{array}{ccc}
          0 & 0 & 0 \\
          0 & 2g\Delta & 0 \\
          0 & 0 & 0
          \end{array}\right) \ \ , \ \ 
\eta_{-}=\Box^{1/2}P_{-}\theta^{2}P_{-} \ \ , \ \ 
\bar{\eta}_{+}=\Box^{1/2}P_{+}\bar{\theta}^{2}P_{+} \ , \label{matrizes}                   
\end{equation}
is the quadratic operator of the free part of the Lagrangian.

The basic techniques for the calculation of (\ref{order 1}) for general supersymmetric models have been developed in \cite{Nibbelink}, and have been applied to the O'Raifeartaigh model in the context of the linear delta expansion in \cite{paper 2}. So, we refer the reader to these references for extensive details.

The one-loop diagram is given by \cite{paper 2}
\begin{eqnarray}
{\mathcal V}_{eff}^{(1)}&=&-\frac{1}{2}tr\left[L_0(\tilde{A}+\tilde{B})+L_0(\tilde{A}-\tilde{B})-2L_0(\tilde{A})\right] \nonumber\\
&=&-\frac{1}{(8\pi)^{2}}\left\{(m^{2}+4g^{2}\langle z_{1}\rangle^{2})^{2}\ln\left[1-\frac{4g^{2}\langle h_{0}\rangle^{2}}{(m^{2}+4g^{2}\langle z_{1}\rangle^{2})^{2}}\right]\right. \hspace{1,5cm}\nonumber\\
&&\left.+4g\langle h_{0}\rangle(m^{2}+4g^{2}\langle z_{1}\rangle^{2})\ln\frac{m^{2}+4g^{2}\langle z_{1}\rangle^{2}+2g\langle h_{0}\rangle}{m^{2}+4g^{2}\langle z_{1}\rangle^{2}-2g\langle h_{0}\rangle}\right. \nonumber\\
&&\left.+4g^{2}\langle h_{0}\rangle^{2}\ln\left[(m^{2}+4g^{2}\langle z_{1}\rangle^{2})^{2}-4g^{2}\langle h_{0}\rangle^{2}\right]\right\} \ , \label{V1}
\end{eqnarray}
where $\tilde{A}=A\bar{A}$, $\tilde{B}=\left(B\bar{B}\right)^{1/2}$, and
\begin{equation}
L_0(X)=\int\frac{d^{4}k}{(2\pi)^{4}}\ln\left(1+\frac{X}{k^2}\right) \ .
\end{equation}


\section{The two-loop effective potential}


In this section, we are going to calculate the contribution of the two-loop vacuum diagrams in superspace. As pointed out in the previous section, the propagators have a nontrivial dependence on the spurion insertions. However, using the spurion algebraic relations described in the Appendix, we can solve the $\theta$- and $\bar{\theta}$-dependent integrals and the remaining ones are usual momentum-space loop integrals. We choose to carry out our calculations with renormalized parameters and we are going to adopt a usual renormalization procedure in the next section. 

The two-loop diagrams we have to calculate are shown in Fig. 2. 

\begin{eqnarray*}
\begin{picture}(270,10) \thicklines 
\put(15,0){\circle{40}}\put(15,-20){\line(0,50){40}}\put(0,24){$\phi_0$}\put(0,-28){$\phi_0$}\put(17,24){$\phi_1$}\put(17,-28){$\phi_1$}\put(3,-12){$\phi_1$}\put(16,9){$\phi_1$}\put(38,-3){+}\put(52,-3){$h.c.$}\put(73,-3){+}
\put(108,0){\circle{40}}\put(108,-20){\line(0,50){40}}\put(94,24){$\phi_0$}\put(94,-29){$\phi_1$}\put(111,24){$\phi_1$}\put(111,-29){$\phi_0$}\put(96,-12){$\phi_1$}\put(109,9){$\phi_1$}\put(131,-3){+}\put(145,-3){$h.c.$}\put(166,-3){+}
\put(203,0){\circle{40}}\put(203,-20){\line(0,50){40}}\put(188,24){$\bar{\phi}_0$}\put(188,-29){$\phi_1$}\put(205,24){$\bar{\phi}_1$}\put(207,-29){$\phi_1$}\put(191,-12){$\phi_0$}\put(204,7){$\bar{\phi}_1$}\put(228,-3){+}
\put(265,0){\circle{40}}\put(265,-20){\line(0,50){40}}\put(250,24){$\bar{\phi}_1$}\put(250,-29){$\phi_1$}\put(267,24){$\bar{\phi}_1$}\put(269,-29){$\phi_1$}\put(253,-12){$\phi_0$}\put(266,7){$\bar{\phi}_0$}
\end{picture}
\end{eqnarray*}

\vspace{0,5cm}
\begin{center}
Fig. 2: Two-loop vacuum diagrams. 
\end{center}
Note that there is only one kind of topology, since in superspace there is only one kind of interaction. Denoting $a=2gx$ and $b=2g\Delta$ for simplicity, the contribution of the first diagram is given by
\begin{eqnarray}
I_1&=&(\!-\!2g)(\!-\!2g)\!\!\int\!\frac{d^{4}pd^{4}kd^{4}\theta_{12}}{(2\pi)^8}\!\left[\!-\frac{1}{4}\bar{D}_{1}^{2}(p)\langle\phi_{0}\phi_{0}\rangle\right]\!\left[\!-\frac{1}{4}\bar{D}_{2}^{2}(k)\langle\phi_{1}\phi_{1}\rangle\right]\!\left[\!\frac{1}{16}\bar{D}_{1}^{2}(q)\bar{D}_{2}^{2}(-q)\langle\phi_{1}\phi_{1}\rangle\right] \nonumber\\
&=&-\frac{g^{2}}{(16)^{3}}a^{2}b^{3}\int\!\frac{d^{4}pd^{4}k}{(2\pi)^8}C(p)F(k)F(q){\mathcal I}_{1}(\theta,\bar\theta) \ , \label{I1}
\end{eqnarray}
where $q=k-p$ and
\begin{eqnarray}
{\mathcal I}_{1}(\theta,\bar\theta)&=&\int\!d^{4}\theta_{12}\left[\bar{D}_{1}^{2}(p)D_{1}^{2}(p)\theta_{1}^{2}\delta_{12}^{4}\right]\left[\bar{D}_{2}^{2}(k)\bar{\theta}_{2}^{2}D_{2}^{2}(k)\delta_{12}^{4}\right]\left[\bar{D}_{1}^{2}(q)\bar{D}_{2}^{2}(-q)\bar{\theta}_{1}^{2}D_{1}^{2}(q)\delta_{12}^{4}\right] \nonumber\\
&=&4(16)^{3}p^{2} \ . \label{I1theta}
\end{eqnarray}
Here, we have used the algebra of covariant derivatives and the spurion algebraic relations described in the Appendix. The same sort of algebraic manipulations is going to be carried out in the sequel to perform the superspace integrals. In the Appendix, we explicitly calculate two superspace integrals as examples of these manipulations.

Plugging this result into (\ref{I1}), we obtain
\begin{equation}
I_1=-4g^{2}a^{2}b^{3}\int\!\frac{d^{4}pd^{4}k}{(2\pi)^8}C(p)F(k)F(q)p^{2} \ . \label{diagram 1}
\end{equation}

The contribution of the second diagram is given by
\begin{eqnarray}
I_2&=&2(\!-\!2g)(\!-\!2g)\!\!\int\!\frac{d^{4}pd^{4}kd^{4}\theta_{12}}{(2\pi)^8}\!\!\left[\!-\frac{1}{4}\bar{D}_{1}^{2}(p)\langle\phi_{0}\phi_{1}\rangle\right]\!\!\left[\!-\frac{1}{4}\bar{D}_{2}^{2}(k)\langle\phi_{0}\phi_{1}\rangle\right]\!\!\left[\!\frac{1}{16}\bar{D}_{1}^{2}(q)\bar{D}_{2}^{2}(-q)\langle\phi_{1}\phi_{1}\rangle\right] \nonumber\\
&=&\frac{2g^{2}}{(16)^{3}}a^{2}b\int\!\frac{d^{4}pd^{4}k}{(2\pi)^8}\left\{A(p)A(k)F(q){\mathcal I}_{2}(\theta,\bar\theta)-2b^{2}A(p)B(k)F(q){\mathcal I}_{3}(\theta,\bar\theta)\right. \nonumber\\
&& \ \ \ \ \ \ \ \ \ \ \ \ \ \ \ \ \ \ \ \ \ \ \ \ \ \ \left.+b^{4}B(p)B(k)F(q){\mathcal I}_{4}(\theta,\bar\theta)\right\} \ , \label{I2}
\end{eqnarray}
where
\begin{eqnarray}
{\mathcal I}_{2}(\theta,\bar\theta)&=&\int\!d^{4}\theta_{12}\left[\bar{D}_{1}^{2}(p)D_{1}^{2}(p)\delta_{12}^{4}\right]\left[\bar{D}_{2}^{2}(k)D_{2}^{2}(k)\delta_{12}^{4}\right]\left[\bar{D}_{1}^{2}(q)\bar{D}_{2}^{2}(-q)\bar{\theta}_{1}^{2}D_{1}^{2}(q)\delta_{12}^{4}\right] \nonumber\\
&=&0 \ , \label{I2theta}
\end{eqnarray}
\begin{eqnarray}
{\mathcal I}_{3}(\theta,\bar\theta)&=&\int\!d^{4}\theta_{12}\left[\bar{D}_{1}^{2}(p)D_{1}^{2}(p)\delta_{12}^{4}\right]\left[\bar{D}_{2}^{2}(k)\theta_{2}^{2}\bar{\theta}_{2}^{2}D_{2}^{2}(k)\delta_{12}^{4}\right]\left[\bar{D}_{1}^{2}(q)\bar{D}_{2}^{2}(-q)\bar{\theta}_{1}^{2}D_{1}^{2}(q)\delta_{12}^{4}\right] \nonumber\\
&=&4(16)^{3}p^{2} \ , \label{I3theta}
\end{eqnarray}
\begin{eqnarray}
{\mathcal I}_{4}(\theta,\bar\theta)&=&\int\!d^{4}\theta_{12}\left[\bar{D}_{1}^{2}(p)\theta_{1}^{2}\bar{\theta}_{1}^{2}D_{1}^{2}(p)\delta_{12}^{4}\right]\left[\bar{D}_{2}^{2}(k)\theta_{2}^{2}\bar{\theta}_{2}^{2}D_{2}^{2}(k)\delta_{12}^{4}\right]\left[\bar{D}_{1}^{2}(q)\bar{D}_{2}^{2}(-q)\bar{\theta}_{1}^{2}D_{1}^{2}(q)\delta_{12}^{4}\right] \nonumber\\
&=&-4(16)^{3} \ . \label{I4theta}
\end{eqnarray}

With (\ref{I2theta})-(\ref{I4theta}) into (\ref{I2}), we get the contribution:
\begin{eqnarray}
I_2=-16g^{2}a^{2}b^{3}\int\!\frac{d^{4}pd^{4}k}{(2\pi)^8}A(p)B(k)F(q)p^{2}-8g^{2}a^{2}b^{5}\int\!\frac{d^{4}pd^{4}k}{(2\pi)^8}B(p)B(k)F(q) \ . \label{diagram 2}
\end{eqnarray}

The contribution of the third diagram is given by
\begin{eqnarray}
I_3&=&4(\!-\!2g)(\!-\!2g)\!\!\int\!\frac{d^{4}pd^{4}kd^{4}\theta_{12}}{(2\pi)^8}\!\!\left[\!-\frac{1}{4}\bar{D}_{1}^{2}(p)\langle\phi_{1}\bar{\phi}_{0}\rangle\right]\!\!\left[\!-\frac{1}{4}D_{2}^{2}(k)\langle\bar{\phi}_{1}\phi_{1}\rangle\right]\!\!\left[\!\frac{1}{16}\bar{D}_{1}^{2}(q)D_{2}^{2}(-q)\langle\phi_{0}\bar{\phi}_{1}\rangle\right] \nonumber\\
&=&\frac{g^{2}}{(16)^{3}}a^{2}b^{2}\int\!\frac{d^{4}pd^{4}k}{(2\pi)^8}C(p)C(q)\left\{E(k){\mathcal I}_{5}(\theta,\bar\theta)+\frac{1}{16}b^{2}B(k){\mathcal I}_{6}(\theta,\bar\theta)\right\} \ , \label{I3}
\end{eqnarray}
where
\begin{eqnarray}
{\mathcal I}_{5}(\theta,\bar\theta)&=&\int\!d^{4}\theta_{12}\left[\bar{D}_{1}^{2}(p)\bar{\theta}_{1}^{2}D_{1}^{2}(p)\bar{D}_{1}^{2}(p)\delta_{12}^{4}\right]\left[D_{2}^{2}(k)\delta_{12}^{4}\right]\left[\bar{D}_{1}^{2}(q)D_{2}^{2}(-q)D_{1}^{2}(q)\bar{D}_{1}^{2}(q)\theta_{1}^{2}\delta_{12}^{4}\right] \nonumber\\
&=&(16)^{4}p^{2}q^{2} \ , \label{I5theta}
\end{eqnarray}
\begin{eqnarray}
{\mathcal I}_{6}(\theta,\bar\theta)&=&\int\!d^{4}\theta_{12}\left[\bar{D}_{1}^{2}(p)\bar{\theta}_{1}^{2}D_{1}^{2}(p)\bar{D}_{1}^{2}(p)\delta_{12}^{4}\right]\left[D_{2}^{2}(k)\bar{D}_{2}^{2}(k)\bar{\theta}_{2}^{2}\theta_{2}^{2}D_{2}^{2}(k)\delta_{12}^{4}\right] \nonumber\\
&& \ \ \ \ \ \ \ \times \left[\bar{D}_{1}^{2}(q)D_{2}^{2}(-q)D_{1}^{2}(q)\bar{D}_{1}^{2}(q)\theta_{1}^{2}\delta_{12}^{4}\right] \nonumber\\
&=&(16)^{5}p^{2}q^{2} \ . \label{I6theta}
\end{eqnarray}

By inserting (\ref{I5theta}) and (\ref{I6theta}) into (\ref{I3}), we obtain
\begin{equation}
I_3=16g^{2}a^{2}b^{2}\int\!\frac{d^{4}pd^{4}k}{(2\pi)^8}C(p)E(k)C(q)p^{2}q^{2}+16g^{2}a^{2}b^{4}\int\!\frac{d^{4}pd^{4}k}{(2\pi)^8}C(p)B(k)C(q)p^{2}q^{2} \ . \label{diagram 3}
\end{equation}

The contribution of the fourth diagram is given by
\begin{eqnarray}
I_4&=&2(\!-\!2g)(\!-\!2g)\!\!\int\!\frac{d^{4}pd^{4}kd^{4}\theta_{12}}{(2\pi)^8}\!\!\left[\!-\frac{1}{4}\bar{D}_{1}^{2}(p)\langle\phi_{1}\bar{\phi}_{1}\rangle\right]\!\!\left[\!-\frac{1}{4}D_{2}^{2}(k)\langle\bar{\phi}_{1}\phi_{1}\rangle\right]\!\!\left[\!\frac{1}{16}\bar{D}_{1}^{2}(q)D_{2}^{2}(-q)\langle\phi_{0}\bar{\phi}_{0}\rangle\right] \nonumber\\
&=&\frac{8g^{2}}{(16)^{2}}a^{2}b^{2}\int\!\frac{d^{4}pd^{4}k}{(2\pi)^8}B(q)\left\{E(p)E(k){\mathcal I}_{7}(\theta,\bar\theta)+\frac{1}{8}b^{2}E(p)B(k){\mathcal I}_{8}(\theta,\bar\theta)\right. \nonumber\\
&& \ \ \ \ \ \ \ \ \ \ \ \ \ \ \ \ \ \ \ \ \ \ \ \ \ \ \ \ \ \ \ \ \ \ \left.+\frac{1}{(16)^{2}}b^{4}B(p)B(k){\mathcal I}_{9}(\theta,\bar\theta)\right\} \nonumber\\
&&+\frac{8g^{2}}{(16)^{2}}\int\!\frac{d^{4}pd^{4}k}{(2\pi)^8}A(q)(q^{2}+m^{2})\left\{E(p)E(k){\mathcal I}_{10}(\theta,\bar\theta)+\frac{1}{8}b^{2}E(p)B(k){\mathcal I}_{11}(\theta,\bar\theta)\right. \nonumber\\
&& \ \ \ \ \ \ \ \ \ \ \ \ \ \ \ \ \ \ \ \ \ \ \ \ \ \ \ \ \ \ \ \ \ \ \ \ \ \ \ \ \ \ \ \ \left.+\frac{1}{(16)^{2}}b^{4}B(p)B(k){\mathcal I}_{12}(\theta,\bar\theta)\right\} \ , \label{I4}
\end{eqnarray}
where
\begin{eqnarray}
{\mathcal I}_{7}(\theta,\bar\theta)&=&\int\!d^{4}\theta_{12}\left[\bar{D}_{1}^{2}(p)\delta_{12}^{4}\right]\left[D_{2}^{2}(k)\delta_{12}^{4}\right]\left[\bar{D}_{1}^{2}(q)D_{2}^{2}(-q)\theta_{1}^{2}\bar{\theta}_{1}^{2}\delta_{12}^{4}\right] \nonumber\\
&=&(16)^{2} \ , \label{I7theta}
\end{eqnarray}
\begin{eqnarray}
{\mathcal I}_{8}(\theta,\bar\theta)&=&\int\!d^{4}\theta_{12}\left[\bar{D}_{1}^{2}(p)\delta_{12}^{4}\right]\left[D_{2}^{2}(k)\bar{D}_{2}^{2}(k)\bar{\theta}_{2}^{2}\theta_{2}^{2}D_{2}^{2}(k)\delta_{12}^{4}\right]\left[\bar{D}_{1}^{2}(q)D_{2}^{2}(-q)\theta_{1}^{2}\bar{\theta}_{1}^{2}\delta_{12}^{4}\right] \nonumber\\
&=&(16)^{3} \ , \label{I8theta}
\end{eqnarray}
\begin{eqnarray}
{\mathcal I}_{9}(\theta,\bar\theta)&=&\int\!d^{4}\theta_{12}\left[\bar{D}_{1}^{2}(p)D_{1}^{2}(p)\theta_{1}^{2}\bar{\theta}_{1}^{2}\bar{D}_{1}^{2}(p)\delta_{12}^{4}\right]\left[D_{2}^{2}(k)\bar{D}_{2}^{2}(k)\bar{\theta}_{2}^{2}\theta_{2}^{2}D_{2}^{2}(k)\delta_{12}^{4}\right] \nonumber\\
&& \ \ \ \ \ \ \ \times \left[\bar{D}_{1}^{2}(q)D_{2}^{2}(-q)\theta_{1}^{2}\bar{\theta}_{1}^{2}\delta_{12}^{4}\right] \nonumber\\
&=&(16)^{4} \ , \label{I9theta}
\end{eqnarray}
\begin{eqnarray}
{\mathcal I}_{10}(\theta,\bar\theta)&=&\int\!d^{4}\theta_{12}\left[\bar{D}_{1}^{2}(p)\delta_{12}^{4}\right]\left[D_{2}^{2}(k)\delta_{12}^{4}\right]\left[\bar{D}_{1}^{2}(q)D_{2}^{2}(-q)\delta_{12}^{4}\right] \nonumber\\
&=&0 \ , \label{I10theta}
\end{eqnarray}
\begin{eqnarray}
{\mathcal I}_{11}(\theta,\bar\theta)&=&\int\!d^{4}\theta_{12}\left[\bar{D}_{1}^{2}(p)\delta_{12}^{4}\right]\left[D_{2}^{2}(k)\bar{D}_{2}^{2}(k)\bar{\theta}_{2}^{2}\theta_{2}^{2}D_{2}^{2}(k)\delta_{12}^{4}\right]\left[\bar{D}_{1}^{2}(q)D_{2}^{2}(-q)\delta_{12}^{4}\right] \nonumber\\
&=&-(16)^{3}k^{2} \ , \label{I11theta}
\end{eqnarray}
\begin{eqnarray}
{\mathcal I}_{12}(\theta,\bar\theta)&=&\int\!d^{4}\theta_{12}\left[\bar{D}_{1}^{2}(p)D_{1}^{2}(p)\theta_{1}^{2}\bar{\theta}_{1}^{2}\bar{D}_{1}^{2}(p)\delta_{12}^{4}\right]\left[D_{2}^{2}(k)\bar{D}_{2}^{2}(k)\bar{\theta}_{2}^{2}\theta_{2}^{2}D_{2}^{2}(k)\delta_{12}^{4}\right] \nonumber\\
&& \ \ \ \ \ \ \ \times \left[\bar{D}_{1}^{2}(q)D_{2}^{2}(-q)\delta_{12}^{4}\right] \nonumber\\
&=&-(16)^{4}q^{2} \ . \label{I12theta}
\end{eqnarray}

By taking (\ref{I7theta})-(\ref{I12theta}) into (\ref{I4}), the final contribution reads as follows:
\begin{eqnarray}
I_4&=&8g^{2}a^{2}b^{2}\int\!\frac{d^{4}pd^{4}k}{(2\pi)^8}E(p)E(k)B(q)+16g^{2}a^{2}b^{4}\int\!\frac{d^{4}pd^{4}k}{(2\pi)^8}E(p)B(k)B(q) \nonumber\\
&&+8g^{2}a^{2}b^{6}\int\!\frac{d^{4}pd^{4}k}{(2\pi)^8}B(p)B(k)B(q) \nonumber\\
&&-16g^{2}b^{2}\int\!\frac{d^{4}pd^{4}k}{(2\pi)^8}E(p)B(k)A(q)k^{2}q^{2}-8g^{2}b^{4}\int\!\frac{d^{4}pd^{4}k}{(2\pi)^8}B(p)B(k)A(q)q^{4} \nonumber\\
&&-16g^{2}m^{2}b^{2}\int\!\frac{d^{4}pd^{4}k}{(2\pi)^8}E(p)B(k)A(q)k^{2}-8g^{2}m^{2}b^{4}\int\!\frac{d^{4}pd^{4}k}{(2\pi)^8}B(p)B(k)A(q)q^{2} \ . \label{diagram 4}
\end{eqnarray}

Once the superspace sector of $I_1$, $I_2$, $I_3$, and $I_4$ has been worked out, we are ready to compute the momentum-space two-loop integrals to get the two-loop corrected effective potential we are looking for.


\section{The two-loop contribution to the effective potential}

In this section, we give the expressions for the two-loop vacuum diagrams in terms of integrals in momentum space. The explicit derivation of these expressions is reported in Appendix B.

In Sec. II, we have used the spurion algebraic relations to reduce the superspace integrals to usual integrals over the momenta of the loops. It can be readily checked, by power counting, that some of the integrals are finite while some of them are log divergent. To handle these integrals, we have adopted the following strategy: for each of them, we split the integrand with the help of the method of partial fraction decomposition and write each integral as the sum of other integrals with just three terms in the denominator. The remaining integrals are all well known in the literature, and we use the results of the references \cite{Jones,Espinosa,Martin} to compute them. 

From now on, we define $\eta^{2}=m^{2}+a^{2}$, $\eta^{\pm}=m^{2}+a^{2}\pm b$ and adopt the same notation of references \cite{Jones,Martin} for the integrals $I(x,y,z)$ , $J(x,y)$, and $J(x)$:
\begin{equation}
\kappa J(x)=-\frac{x}{\epsilon}+x\left(\bar{\ln}x-1\right) \ , \label{J(x)}
\end{equation}
\begin{equation}
\kappa^{2}J(x,y)=xy\left[\frac{1}{\epsilon^{2}}+\frac{1}{\epsilon}\left(2-\bar{\ln}x-\bar{\ln}y\right)+\left(1-\bar{\ln}x-\bar{\ln}y+\bar{\ln}x\bar{\ln}y\right)\right] \ , \label{J(x,y)}
\end{equation}
\begin{eqnarray}
\kappa^{2}I(x,y,z)&=&-\frac{c}{2\epsilon^2}-\frac{1}{\epsilon}\left(\frac{3c}{2}-L_1\right)-\frac{1}{2}\left\{L_2-6L_1+(y\!+\!z\!-\!x)\bar{\ln}y\bar{\ln}z\right. \nonumber\\
&&\left.+(z\!+\!x\!-\!y)\bar{\ln}z\bar{\ln}x+(y\!+\!x\!-\!z)\bar{\ln}y\bar{\ln}x+\xi(x,y,z)+c\left[7+\zeta(2)\right]\right\} \ , \label{I(x,y,z)}
\end{eqnarray}
where
\begin{eqnarray*}
\kappa&=&(4\pi)^{2} \ , \\
c&=&x+y+z \ , \\
\bar{\ln}X&=&\ln\left(\frac{X}{\mu^{2}}\right)+\gamma-\ln4\pi \ , \\
L_m&=&x\bar{\ln}^{m}x+y\bar{\ln}^{m}y+z\bar{\ln}^{m}z \ , \\
\xi(x,y,z)&=&S\left[2\ln\frac{z+x-y-S}{2z}\ln\frac{z+y-x-S}{2z}-\ln\frac{x}{z}\ln\frac{y}{z}\right. \\
&&\left. \ \ \ \ \ -2\mbox{Li}_{2}\left(\frac{z+x-y-S}{2z}\right)-2\mbox{Li}_{2}\left(\frac{z+y-x-S}{2z}\right)+\frac{\pi^2}{3}\right] \ , \\
S&=&\sqrt{x^2+y^2+z^2-2xy-2yz-2zx} \ , \\
\mbox{Li}_{2}(z)&=&-\int_{0}^{z}\frac{\ln(1-t)}{t}dt \ \ (\mbox{dilogarithm function}) \ .
\end{eqnarray*}

Below, we cast the final expressions for the two-loop diagrams.

For the first two-loop diagram, Eq. (\ref{diagram 1}), we have
\begin{equation}
I_1=\frac{g^{2}a^{2}}{2}\left[I(\eta^{+},\eta^{+},\eta^{+})\!-\!3I(\eta^{+},\eta^{+},\eta^{-})\!+\!3I(\eta^{+},\eta^{-},\eta^{-})\!-\!I(\eta^{-},\eta^{-},\eta^{-})\right] \ . \label{diagram 1 integrals}
\end{equation}

For the second two-loop diagram, Eq. (\ref{diagram 2}):
\begin{eqnarray}
I_2&=&g^{2}a^{2}\left[-4I(\eta^{2},\eta^{2},\eta^{+})+4I(\eta^{2},\eta^{2},\eta^{-})+I(\eta^{+},\eta^{+},\eta^{+})+I(\eta^{+},\eta^{+},\eta^{-})\right. \nonumber\\
&& \ \ \ \ \ \ \ \left.-I(\eta^{+},\eta^{-},\eta^{-})-I(\eta^{-},\eta^{-},\eta^{-})\right] \ . \label{diagram 2 integrals}
\end{eqnarray}

For the third two-loop diagram, Eq. (\ref{diagram 3}):
\begin{equation}
I_3=2g^{2}a^{2}\left[I(\eta^{+},\eta^{+},\eta^{+})-I(\eta^{+},\eta^{+},\eta^{-})-I(\eta^{+},\eta^{-},\eta^{-})+I(\eta^{-},\eta^{-},\eta^{-})\right] \ . \label{diagram 3 integrals}
\end{equation}

For the fourth two-loop diagram, Eq. (\ref{diagram 4}):
\begin{eqnarray}
I_4&=&8g^{2}a^{2}\left[-I(\eta^{2},\eta^{2},\eta^{2})+\frac{b}{\eta^{2}}I(\eta^{2},\eta^{2},\eta^{+})-\frac{b}{\eta^{2}}I(\eta^{2},\eta^{2},\eta^{-})\right] \nonumber\\
&&+8g^{2}m^{2}\left[-2I(\eta^{2},\eta^{2},0)+\frac{\eta^+}{\eta^2}I(\eta^{2},\eta^{+},0)+\frac{\eta^-}{\eta^2}I(\eta^{2},\eta^{-},0)\right] \nonumber\\
&&+g^{2}a^{2}\left[I(\eta^{+},\eta^{+},\eta^{+})+3I(\eta^{+},\eta^{+},\eta^{-})+3I(\eta^{+},\eta^{-},\eta^{-})+I(\eta^{-},\eta^{-},\eta^{-})\right] \nonumber\\
&&+2g^{2}\left[-4J(\eta^{2}\!,\!\eta^{2})\!+\!4J(\eta^{2}\!,\!\eta^{+})\!+\!4J(\eta^{2}\!,\!\eta^{-})\!-\!J(\eta^{+}\!,\!\eta^{+})\!-\!2J(\eta^{+}\!,\!\eta^{-})\!-\!J(\eta^{-}\!,\!\eta^{-})\right] \ . \label{diagram 4 integrals}
\end{eqnarray}

Now, the renormalized two-loop effective potential is finally given by the finite part of (\ref{V2 integrals}) derived in Appendix B, and can be written as
\begin{eqnarray}
{\mathcal V}_{eff}^{(2)r}&=&\frac{g^{2}}{(4\pi)^{4}}\left\{\frac{4}{m^{2}+a^{2}}\left[(m^{2}+a^{2})^{2}(2m^{2}+a^{2})+b(m^{2}+a^{2}-b)(m^{2}-2a^{2})\right]\bar{\ln}^{2}(m^{2}+a^{2})\right. \nonumber\\
&&\left.+\frac{(m^{2}+a^{2}+b)}{(m^{2}+a^{2})}\left[4m^{2}b-(m^{2}+a^{2})(2m^{2}+11a^{2}+2b)\right]\bar{\ln}^{2}(m^{2}+a^{2}+b)\right. \nonumber\\
&&\left.+\left[b(3a^{2}-2b)-(m^{2}+a^{2})(2m^{2}+3a^{2}-4b)\right]\bar{\ln}^{2}(m^{2}+a^{2}-b)\right. \nonumber\\
&&\left.+8(m^{2}+a^{2}+b)(2a^{2}-b)\bar{\ln}(m^{2}+a^{2})\bar{\ln}(m^{2}+a^{2}+b)\right. \nonumber\\
&&\left.+\frac{8a^{2}b(m^{2}+a^{2}-b)}{(m^{2}+a^{2})}\bar{\ln}(m^{2}+a^{2})\bar{\ln}(m^{2}+a^{2}-b)\right. \nonumber\\
&&\left.-2(m^{2}+a^{2}+b)(2m^{2}+3a^{2}-2b)\bar{\ln}(m^{2}+a^{2}+b)\bar{\ln}(m^{2}+a^{2}-b)\right. \nonumber\\
&&\left.+\frac{8m^{2}b}{m^{2}\!+\!a^{2}}\!\left[2b\!\ln(m^{2}\!+\!a^{2})\!\!-\!\!(m^{2}\!+\!a^{2}\!+\!b)\!\ln(m^{2}\!+\!a^{2}\!+\!b)\!\!+\!\!(m^{2}\!+\!a^{2}\!-\!b)\!\ln(m^{2}\!+\!a^{2}\!-\!b)\right]\!\bar{\ln}b\right. \nonumber\\
&&\left.-16(m^{2}+a^{2})(2m^{2}+3a^{2})\bar{\ln}(m^{2}+a^{2})\right. \nonumber\\
&&\left.+8(m^{2}+a^{2}+b)(2m^{2}+3a^{2}+2b)\bar{\ln}(m^{2}+a^{2}+b)\right. \nonumber\\
&&\left.+8(m^{2}+a^{2}-b)(2m^{2}+3a^{2}-2b)\bar{\ln}(m^{2}+a^{2}-b)\right. \nonumber\\
&&\left.+a^{2}\!\left[4\xi(m^{2}\!+\!a^{2},m^{2}\!+\!a^{2},m^{2}\!+\!a^{2})\!-\!3\xi(m^{2}\!+\!a^{2}\!+\!b,m^{2}\!+\!a^{2}\!+\!b,m^{2}\!+\!a^{2}\!+\!b)\right.\right. \nonumber\\
&&\left.\left. \ \ \ \ \ \ -\xi(m^{2}\!+\!a^{2}\!+\!b,m^{2}\!+\!a^{2}\!-\!b,m^{2}\!+\!a^{2}\!-\!b)\right]\right. \nonumber\\
&&\left.+\frac{4(m^{2}+a^{2}-b)}{m^{2}+a^{2}}\left[\xi(m^{2}\!+\!a^{2},m^{2}\!+\!a^{2},m^{2}\!+\!a^{2}\!+\!b)-\xi(m^{2}\!+\!a^{2},m^{2}\!+\!a^{2},m^{2}\!+\!a^{2}\!-\!b)\right]\right. \nonumber\\
&&\left.+\frac{8m^{2}b}{m^{2}+a^{2}}\left[(m^{2}+a^{2}+b)\mbox{Li}_{2}\!\left(\frac{m^{2}+a^{2}}{m^{2}+a^{2}+b}\right)+(m^{2}+a^{2}-b)\mbox{Li}_{2}\!\left(\frac{m^{2}+a^{2}-b}{m^{2}+a^{2}}\right)\right]\right. \nonumber\\
&&\left.-40b^{2}-\frac{8}{3}m^{2}b\pi^{2}\right\} \ , \label{V2 final}
\end{eqnarray}
where, as before, $a=2gx$ and $b=2g\Delta$.

Our two-loop expression of Eq.(\ref{V2 final}) for the effective potential as given in component fields is attained by projecting the result of our superspace computation in terms of superfield Feynman rules. A direct component-field two-loop calculation of the effective potential for the O'Raifeartaigh model is not available in the literature. However, in close connection with our result, we quote the work by Miller \cite{Miller}, where the component-field two-loop corrected effective potential of the Wess-Zumino model is derived.

To conclude this section, we would like to comment on the fact that our two-loop result could also be calculated from the wave function renormalizations accounted for in the two-loop corrected K\"{a}hler potential appropriate to describe the O'Raifeartaigh model. The work by Nibbelink and Nyawelo cited in Ref.\cite{Kpotential}, where two-loop effective K\"{a}hler potentials are calculated for supersymmetric models, once applied to the minimal O'Raifeartaigh model and suitably projected into component fields, can also be used to compute the effective potential we have presented in this work.


\section{Concluding Remarks}

Explicit and spontaneous SUSY breakdown are topics of relevance in connection with phenomenological and fundamental aspects of particle field theories for the physics of the standard model and the so-called beyond standard model physics. It was our actual goal in this paper to ascertain that, despite SUSY breaking, superspace and superfield techniques are worthy to be kept whenever one wishes to compute higher-order corrections to the effective action and the effective potential.

Summing up all orders in the breaking parameters yield rather nontrivial expressions for the shifted superpropagators. This might, in principle, appear to be a disadvantage to keep on adopting superfield Feynman rules to perform higher-order computations if SUSY is broken. However, the work of Ref. \cite{Helayel} shows an effort to systematize and adapt supergraph techniques even if SUSY is not exact. A number of explicit $\theta$-dependent expressions are written down and the whole procedure of (exact SUSY) supergraph methods is thoroughly extended to account for SUSY explicit and spontaneous breaking.  

We have here devoted our efforts to show the efficacy of the broken-case supergraph procedure in the concrete problem of computing a two-loop corrected effective potential in a way that can be extended to the whole class of O'Raifeartaigh-type and also Fayet-Iliopoulos \cite{FI} (D-term SUSY breaking) models. We succeed in finding shortcuts - and we explicitly show them - which confirm the benefits and efficacy of superspace methods to carry out loop calculations whenever SUSY is no longer exact. In our present case, the supergraph procedure drastically reduces (from hundreds) the number of diagrams to be computed. Superpropagators become however much more cumbersome. Nevertheless, to deal with them is not complicated as it might appear at a first glance, in view of the anticommuting character of the $\theta$ variable and the tricks and special simplifying recursive relations we develop to treat the long $\theta$ expressions that appear throughout the (broken-case) superspace loop evaluations. The option of drawing and calculating very few supergraphs, even if super-Feynman rules get more involved, is favored and we confirm this claim in our explicit two-loop evaluation of the effective potential reported here.  

As we have pointed out in the Introduction, the works of Refs.\cite{IS} and \cite{ISS1,ISS2,Shih,Marques,twoloop} tackle the interesting question of \textit{R}-symmetry spontaneous breakdown and the existence of metastable SUSY breaking vacua. In the context of the Intriligator-Seiberg-Shih mechanism, a natural follow-up of the present work would consist in a more detailed study and discussion of our loop-corrected effective potential to analyze issues like its effects on the lifetime of these unstable vacua for the minimal and nonminimal O'Raifeartaigh-type models. Along this line - in view of the breaking of \textit{R} symmetry - we are also pursuing a more phenomenological investigation, by applying the results reported here to analyze the decay modes of the so-called lightest supersymmetric particle into standard model particles. We shall be reporting on these questions elsewhere \cite{progress}. 


\section{Acknowledgements}

We would like to thank Dr. Alexandre Schmidt for clarifying discussions. M. C. B. Abdalla thanks CNPq, Grant No. 306276/2009-7, J. A. Helay\"{e}l-Neto thanks CNPq, Grant No. 306670/2009-7, and Daniel L. Nedel thanks CNPq, Grant No. 501317/2009-0, for financial support. Carlos R. Senise, Jr. thanks CAPES-Brazil for financial support. 


\section*{Appendix A: Superspace relations and integrals}

In this Appendix, we present some useful relations involving the covariant derivatives in superspace, and use them to explicitly calculate, for the sake of illustration, two of the superspace integrals appearing in Sec. II. Because of the dependence of the propagators on the spurion interactions, some trivial relations appearing in exact SUSY supergraph calculations have to be modified by $\theta$ and $\bar\theta$ insertions in the case of broken SUSY. We follow the same notation as in \cite{WB}.

The covariant derivatives are given by
\begin{equation}
D_{\alpha}(p)=\partial_{\alpha}-\sigma^{\mu}_{\alpha\dot\alpha}\bar{\theta}^{\dot\alpha}p_{\mu} \ , \label{ap1}
\end{equation}
\begin{equation}
\bar{D}_{\dot\alpha}(p)=-\bar{\partial}_{\dot\alpha}+\theta^{\alpha}\sigma^{\mu}_{\alpha\dot\alpha}p_{\mu} \ , \label{ap2}
\end{equation}
and obey the algebra
\begin{equation}
\left\{D_{\alpha}(p),\bar{D}_{\dot\alpha}(k)\right\}=\sigma^{\mu}_{\alpha\dot\alpha}(p+k)_{\mu} \ . \label{ap3}
\end{equation}

From (\ref{ap1}) and (\ref{ap2}), we can also write
\begin{eqnarray}
D^{2}(p)&=&-\partial^{\alpha}\partial_{\alpha}+2\bar{\theta}_{\dot\alpha}\bar{\sigma}^{\mu\dot{\alpha}\alpha}p_{\mu}\partial_{\alpha}+\bar{\theta}^{2}p^{2} \ , \\ \label{ap4}
\bar{D}^{2}(p)&=&-\bar{\partial}_{\dot\alpha}\bar{\partial}^{\dot\alpha}+2\theta^{\alpha}\sigma^{\mu}_{\alpha\dot\alpha}p_{\mu}\bar{\partial}^{\dot\alpha}+\theta^{2}p^{2} \ . \label{ap5}
\end{eqnarray}

Besides (\ref{ap3}), we have the following (anti)commutation relations:
\begin{equation}
\left\{D_{\alpha},\theta_{\beta}\right\}=-\epsilon_{\alpha\beta} \ , \label{ap6}
\end{equation}
\begin{equation}
\left\{D_{\alpha},\bar{\theta}_{\dot\beta}\right\}=0 \ , \label{ap7}
\end{equation}
\begin{equation}
\left[D_{\alpha},\theta^{2}\right]=2\theta_{\alpha} \ , \label{ap8}
\end{equation}
\begin{equation}
\left[D^{2},\theta_{\alpha}\right]=2D_{\alpha} \ , \label{ap9}
\end{equation}
\begin{equation}
\left[D^{2},\theta^{2}\right]=-4+4\theta^{\alpha}D_{\alpha} \ , \label{ap10}
\end{equation}
and analogous relations for $\bar{D} \ .$

When calculating supergraphs in the broken case, some relations involving the covariant derivatives and the fermionic coordinates proved to be very useful. We list below some of them:
\begin{equation}
\delta^{4}_{12}\bar{D}_{1}^{2}D_{1}^{2}\theta_{1}^{2}\delta^{4}_{12}=16\theta_{1}^{2}\delta^{4}_{12} \ , \label{ap11}
\end{equation}
\begin{equation}
\delta^{4}_{12}\bar{D}_{1}^{2}\theta_{1}^{2}\bar{\theta}_{1}^{2}D_{1}^{2}\delta^{4}_{12}=16\theta_{1}^{2}\bar{\theta}_{1}^{2}\delta^{4}_{12} \ , \label{ap12}
\end{equation}
\begin{equation}
\delta^{4}_{12}\bar{D}_{1}^{2}D_{1}^{2}\theta_{1}^{2}\bar{\theta}_{1}^{2}\bar{D}_{1}^{2}D_{1}^{2}\delta^{4}_{12}=(16)^{2}e^{2\theta_{1}\sigma^{\mu}\bar{\theta}_{1}p_{\mu}}\delta^{4}_{12} \ . \label{ap13}
\end{equation}

To show how the relations above apply, we explicitly calculate, in the sequel, the integrals ${\mathcal I}_{6}(\theta,\bar{\theta})$, Eq.(\ref{I6theta}), and ${\mathcal I}_{9}(\theta,\bar{\theta})$, Eq.(\ref{I9theta}).

\subsection*{Explicit calculation of ${\mathcal I}_{6}(\theta,\bar{\theta})$}

The integral ${\mathcal I}_{6}(\theta,\bar{\theta})$ is given by
\begin{eqnarray}
{\mathcal I}_{6}(\theta,\bar\theta)&=&\int\!d^{4}\theta_{12}\left[\bar{D}_{1}^{2}(p)\bar{\theta}_{1}^{2}D_{1}^{2}(p)\bar{D}_{1}^{2}(p)\delta_{12}^{4}\right]\left[D_{2}^{2}(k)\bar{D}_{2}^{2}(k)\bar{\theta}_{2}^{2}\theta_{2}^{2}D_{2}^{2}(k)\delta_{12}^{4}\right] \nonumber\\
&& \ \ \ \ \ \ \ \times \left[\bar{D}_{1}^{2}(q)D_{2}^{2}(-q)D_{1}^{2}(q)\bar{D}_{1}^{2}(q)\theta_{1}^{2}\delta_{12}^{4}\right] \ .
\end{eqnarray}
Using the transfer properties
\begin{equation}
D_{2}^{2}(k)\delta^{4}_{12}=D_{1}^{2}(-k)\delta^{4}_{12} \ \ \ , \ \ \ \theta_{2}^{2}\delta^{4}_{12}=\theta_{1}^{2}\delta^{4}_{12} \label{transfer} 
\end{equation}
in the second and third brackets, and the relation 
\begin{equation}
\bar{D}_{1}^{2}(q)D_{1}^{2}(q)\bar{D}_{1}^{2}(q)=-16q^{2}\bar{D}_{1}^{2}(q) \label{3D}
\end{equation}
in the third one, we get
\begin{eqnarray}
{\mathcal I}_{6}(\theta,\bar\theta)&=&-16q^{2}\!\int\!d^{4}\theta_{12}\left[\bar{D}_{1}^{2}(p)\bar{\theta}_{1}^{2}D_{1}^{2}(p)\bar{D}_{1}^{2}(p)\delta_{12}^{4}\right]\left[D_{1}^{2}(-k)\theta_{1}^{2}\bar{\theta}_{1}^{2}\bar{D}_{1}^{2}(-k)D_{1}^{2}(-k)\delta_{12}^{4}\right] \nonumber\\
&& \ \ \ \ \ \ \ \ \ \ \ \ \ \ \ \ \times \left[\bar{D}_{1}^{2}(q)\theta_{1}^{2}D_{1}^{2}(q)\delta_{12}^{4}\right] \ .
\end{eqnarray}
Transferring $\bar{\theta}_{1}^{2}$ from the second bracket into the first, $\theta_{1}^{2}$ from the third bracket into the second, and using relation (\ref{ap10}) (and its complex conjugate), we obtain 
\begin{equation}
{\mathcal I}_{6}(\theta,\bar\theta)=-(16)^{2}q^{2}\!\!\int\!\!d^{4}\theta_{12}\left[\bar{\theta}_{1}^{2}D_{1}^{2}(p)\bar{D}_{1}^{2}(p)\delta_{12}^{4}\right]\!\!\left[\theta_{1}^{2}\bar{D}_{1}^{2}(-k)D_{1}^{2}(-k)\delta_{12}^{4}\right]\!\!\left[\bar{D}_{1}^{2}(q)D_{1}^{2}(q)\delta_{12}^{4}\right] \ . 
\end{equation}
Transferring $\theta_{1}^{2}$ from the second bracket into the first and integrating by parts with respect to $\bar{D}_{1}^{2}$ of the third bracket,
\begin{equation}
{\mathcal I}_{6}(\theta,\bar\theta)=-(16)^{2}q^{2}\!\!\int\!\!d^{4}\theta_{12}\left[\bar{D}_{1}^{2}(p)\theta_{1}^{2}\bar{\theta}_{1}^{2}D_{1}^{2}(p)\bar{D}_{1}^{2}(p)\delta_{12}^{4}\right]\!\!\left[\bar{D}_{1}^{2}(-k)D_{1}^{2}(-k)\delta_{12}^{4}\right]\!\!\left[D_{1}^{2}(q)\delta_{12}^{4}\right] \ . 
\end{equation}
Integrating by parts the $D_{1}^{2}$ in the third bracket, and using the relations
\begin{equation}
\delta^{4}_{12}\bar{D}_{1}^{2}D_{1}^{2}\delta^{4}_{12}=16\delta^{4}_{12} \ , \label{2D1}
\end{equation}
\begin{equation}
\delta^{4}_{12}D_{1\alpha}\bar{D}_{1}^{2}D_{1}^{2}\delta^{4}_{12}=0 \ \ \ , \ \ \ \delta^{4}_{12}D_{1}^{2}\bar{D}_{1}^{2}D_{1}^{2}\delta^{4}_{12}=0 \ , \label{2D2}
\end{equation}
we obtain
\begin{equation}
{\mathcal I}_{6}(\theta,\bar\theta)=-(16)^{3}q^{2}\!\int\!d^{4}\theta_{12}\left[D_{1}^{2}(p)\bar{D}_{1}^{2}(p)\theta_{1}^{2}\bar{\theta}_{1}^{2}D_{1}^{2}(p)\bar{D}_{1}^{2}(p)\delta_{12}^{4}\right]\delta_{12}^{4} \ . 
\end{equation}
Using (\ref{ap13}),
\begin{equation}
{\mathcal I}_{6}(\theta,\bar\theta)=-(16)^{5}q^{2}\!\int\!d^{4}\theta e^{2\theta\sigma^{\mu}\bar{\theta}p_{\mu}} \ .
\end{equation}
Recalling that
\begin{equation}
\int\!d^{4}\theta e^{2\theta\sigma^{\mu}\bar{\theta}p_{\mu}}=\int\!d^{4}\theta\left(1+2\theta\sigma^{\mu}\bar{\theta}p_{\mu}-\theta^{2}\bar{\theta}^{2}p^{2}\right)=-p^{2} \ ,
\end{equation}
we finally obtain
\begin{equation}
{\mathcal I}_{6}(\theta,\bar\theta)=(16)^{5}p^{2}q^{2} \ . \label{apI6theta}
\end{equation}

\subsection*{Explicit calculation of ${\mathcal I}_{9}(\theta,\bar{\theta})$}

The integral ${\mathcal I}_{9}(\theta,\bar{\theta})$ is given by
\begin{eqnarray}
{\mathcal I}_{9}(\theta,\bar\theta)&=&\int\!d^{4}\theta_{12}\left[\bar{D}_{1}^{2}(p)D_{1}^{2}(p)\theta_{1}^{2}\bar{\theta}_{1}^{2}\bar{D}_{1}^{2}(p)\delta_{12}^{4}\right]\left[D_{2}^{2}(k)\bar{D}_{2}^{2}(k)\bar{\theta}_{2}^{2}\theta_{2}^{2}D_{2}^{2}(k)\delta_{12}^{4}\right] \nonumber\\
&& \ \ \ \ \ \ \ \times \left[\bar{D}_{1}^{2}(q)D_{2}^{2}(-q)\theta_{1}^{2}\bar{\theta}_{1}^{2}\delta_{12}^{4}\right] \ .
\end{eqnarray}
Using the transfer properties (\ref{transfer}) in the second and third brackets,
\begin{eqnarray}
{\mathcal I}_{9}(\theta,\bar\theta)&=&\int\!d^{4}\theta_{12}\left[\bar{D}_{1}^{2}(p)\bar{\theta}_{1}^{2}D_{1}^{2}(p)\theta_{1}^{2}\bar{D}_{1}^{2}(p)\delta_{12}^{4}\right]\left[D_{1}^{2}(-k)\theta_{1}^{2}\bar{\theta}_{1}^{2}\bar{D}_{1}^{2}(-k)D_{1}^{2}(-k)\delta_{12}^{4}\right] \nonumber\\
&& \ \ \ \ \ \ \ \times \left[\bar{D}_{1}^{2}(q)\bar{\theta}_{1}^{2}\theta_{1}^{2}D_{1}^{2}(q)\delta_{12}^{4}\right] \ .
\end{eqnarray}
Transferring $\bar{\theta}_{1}^{2}$ from the second bracket into the first and using the complex conjugate of (\ref{ap10}), yields
\begin{eqnarray}
{\mathcal I}_{9}(\theta,\bar\theta)&=&-4\int\!d^{4}\theta_{12}\left[\bar{\theta}_{1}^{2}D_{1}^{2}(p)\theta_{1}^{2}\bar{D}_{1}^{2}(p)\delta_{12}^{4}\right]\left[D_{1}^{2}(-k)\theta_{1}^{2}\bar{D}_{1}^{2}(-k)D_{1}^{2}(-k)\delta_{12}^{4}\right] \nonumber\\
&& \ \ \ \ \ \ \ \ \ \ \ \ \times \left[\bar{D}_{1}^{2}(q)\bar{\theta}_{1}^{2}\theta_{1}^{2}D_{1}^{2}(q)\delta_{12}^{4}\right] \ .
\end{eqnarray}
Transferring $\bar{\theta}_{1}^{2}$ from the first bracket into the third, and using the complex conjugate of (\ref{ap10}) again, we are lead to
\begin{equation}
{\mathcal I}_{9}(\theta,\bar\theta)=16\!\int\!d^{4}\theta_{12}\!\left[D_{1}^{2}(p)\theta_{1}^{2}\bar{D}_{1}^{2}(p)\delta_{12}^{4}\right]\!\left[D_{1}^{2}(-k)\theta_{1}^{2}\bar{D}_{1}^{2}(-k)D_{1}^{2}(-k)\delta_{12}^{4}\right]\!\left[\bar{\theta}_{1}^{2}\theta_{1}^{2}D_{1}^{2}(q)\delta_{12}^{4}\right] \ . 
\end{equation}
Transferring $\theta_{1}^{2}$ from the third bracket into the second and using (\ref{ap10}),
\begin{equation}
{\mathcal I}_{9}(\theta,\bar\theta)=-4(16)\int d^{4}\theta_{12}\left[D_{1}^{2}(p)\theta_{1}^{2}\bar{D}_{1}^{2}(p)\delta_{12}^{4}\right]\left[\theta_{1}^{2}\bar{D}_{1}^{2}(-k)D_{1}^{2}(-k)\delta_{12}^{4}\right]\left[\bar{\theta}_{1}^{2}D_{1}^{2}(q)\delta_{12}^{4}\right] \ . 
\end{equation}
Once more, we transfer $\theta_{1}^{2}$ from the second bracket into the first and use (\ref{ap10}):
\begin{equation}
{\mathcal I}_{9}(\theta,\bar\theta)=(16)^{2}\int d^{4}\theta_{12}\left[\theta_{1}^{2}\bar{D}_{1}^{2}(p)\delta_{12}^{4}\right]\left[\bar{D}_{1}^{2}(-k)D_{1}^{2}(-k)\delta_{12}^{4}\right]\left[\bar{\theta}_{1}^{2}D_{1}^{2}(q)\delta_{12}^{4}\right] \ . 
\end{equation}
Transferring $\theta_{1}^{2}$ from the first bracket into the third, integrating by parts the $\bar{D}_{1}^{2}$ in the first bracket and using (\ref{2D1}), we obtain
\begin{equation}
{\mathcal I}_{9}(\theta,\bar\theta)=(16)^{3}\int d^{4}\theta_{12}\delta^{4}_{12}\left[\bar{D}_{1}^{2}(q)\theta_{1}^{2}\bar{\theta}_{1}^{2}D_{1}^{2}(q)\delta_{12}^{4}\right] \ . 
\end{equation}
Now, using (\ref{ap12}),
\begin{equation}
{\mathcal I}_{9}(\theta,\bar\theta)=(16)^{4}\int d^{4}\theta\theta^{2}\bar{\theta}^{2} \ . 
\end{equation}
Recalling that
\begin{equation}
\int d^{4}\theta\theta^{2}\bar{\theta}^{2}=1 \ , 
\end{equation}
we finally obtain
\begin{equation}
{\mathcal I}_{9}(\theta,\bar\theta)=(16)^{4} \ . \label{apI9theta} 
\end{equation}

We hope these manipulations make clearer the sort of algebra procedure we have adopted to carry out the $\theta$-superspace integrals.


\section*{Appendix B: The momentum-space two-loop integrals}

Recalling (\ref{diagram 1}), we have
\begin{eqnarray}
I_1&=&-4g^{2}a^{2}b^{3}\int\!\frac{d^{4}pd^{4}k}{(2\pi)^8}C(p)F(k)F(q)p^{2} \nonumber\\
&=&-4g^{2}a^{2}b^{3}I_{1}(p,k)\ , \label{first diagram}
\end{eqnarray}
with
\begin{eqnarray}
I_{1}(p,k)&=&\int\!\frac{d^{4}pd^{4}k}{(2\pi)^8}\frac{1}{(p^{2}+\eta^{+})(p^{2}+\eta^{-})(k^{2}+\eta^{+})(k^{2}+\eta^{-})(q^{2}+\eta^{+})(q^{2}+\eta^{-})} \nonumber\\
&=&\frac{1}{(2b)^3}\left[-\!I(\eta^{+},\eta^{+},\eta^{+})\!+\!3I(\eta^{+},\eta^{+},\eta^{-})\!-\!3I(\eta^{+},\eta^{-},\eta^{-})\!+\!I(\eta^{-},\eta^{-},\eta^{-})\right] \ . \label{I1momenta}
\end{eqnarray}
To get (\ref{I1momenta}), we have split the integrand using the strategy described above. Although each partial integral is divergent, using (\ref{I(x,y,z)}), the final result for $I_1$ is finite. This is either the case for $I_2$ and $I_3$. Plugging (\ref{I1momenta}) into (\ref{first diagram}) yields (\ref{diagram 1 integrals}).

For the second two-loop diagram, Eq. (\ref{diagram 2}),
\begin{eqnarray}
I_2&=&-16g^{2}a^{2}b^{3}\int\!\frac{d^{4}pd^{4}k}{(2\pi)^8}A(p)B(k)F(q)p^{2}-8g^{2}a^{2}b^{5}\int\!\frac{d^{4}pd^{4}k}{(2\pi)^8}B(p)B(k)F(q) \nonumber\\
&=&-16g^{2}a^{2}b^{3}I_{2}(p,k)-8g^{2}a^{2}b^{5}I_{3}(p,k) \ , \label{second diagram}
\end{eqnarray}
with
\begin{eqnarray}
I_{2}(p,k)&=&\int\!\frac{d^{4}pd^{4}k}{(2\pi)^8}\frac{1}{(p^{2}+\eta^{2})(k^{2}+\eta^{2})(k^{2}+\eta^{+})(k^{2}+\eta^{-})(q^{2}+\eta^{+})(q^{2}+\eta^{-})} \nonumber\\
&=&\frac{1}{(2b)^3}\left[4I(\eta^{2},\eta^{2},\eta^{+})\!-\!4I(\eta^{2},\eta^{2},\eta^{-})\!-\!2I(\eta^{2},\eta^{+},\eta^{+})\!+\!2I(\eta^{2},\eta^{-},\eta^{-})\right] \ , \label{I2momenta}
\end{eqnarray}
\begin{eqnarray}
I_{3}(p,k)&=&\int\!\frac{d^{4}pd^{4}k}{(2\pi)^8}\frac{1}{(p^{2}\!+\!\eta^{2})(p^{2}\!+\!\eta^{+})(p^{2}\!+\!\eta^{-})(k^{2}\!+\!\eta^{2})(k^{2}\!+\!\eta^{+})(k^{2}\!+\!\eta^{-})(q^{2}\!+\!\eta^{+})(q^{2}\!+\!\eta^{-})} \nonumber\\
&=&\frac{1}{(2b)^5}\left[-16I(\eta^{2},\eta^{2},\eta^{+})\!+\!16I(\eta^{2},\eta^{2},\eta^{-})\!+\!16I(\eta^{2},\eta^{+},\eta^{+})\!-\!16I(\eta^{2},\eta^{-},\eta^{-})\right. \nonumber\\
&& \ \ \ \ \ \ \  \left.-4I(\eta^{+},\eta^{+},\eta^{+})\!-\!4I(\eta^{+},\eta^{+},\eta^{-})\!+\!4I(\eta^{+},\eta^{-},\eta^{-})\!+\!4I(\eta^{-},\eta^{-},\eta^{-})\right] \ . \label{I3momenta}
\end{eqnarray}
Inserting (\ref{I2momenta}) and (\ref{I3momenta}) into (\ref{second diagram}) leads to (\ref{diagram 2 integrals}).

For the third two-loop diagram, Eq. (\ref{diagram 3}),
\begin{eqnarray}
I_3&=&16g^{2}a^{2}b^{2}\int\!\frac{d^{4}pd^{4}k}{(2\pi)^8}C(p)E(k)C(q)p^{2}q^{2}+16g^{2}a^{2}b^{4}\int\!\frac{d^{4}pd^{4}k}{(2\pi)^8}C(p)B(k)C(q)p^{2}q^{2} \nonumber\\
&=&16g^{2}a^{2}b^{2}I_{4}(p,k)+16g^{2}a^{2}b^{4}I_{5}(p,k) \ , \label{third diagram}
\end{eqnarray}
with
\begin{eqnarray}
I_{4}(p,k)&=&\int\!\frac{d^{4}pd^{4}k}{(2\pi)^8}\frac{1}{(p^{2}+\eta^{+})(p^{2}+\eta^{-})(k^{2}+\eta^{2})(q^{2}+\eta^{+})(q^{2}+\eta^{-})} \nonumber\\
&=&\frac{1}{(2b)^2}[I(\eta^{2},\eta^{+},\eta^{+})\!-\!2I(\eta^{2},\eta^{+},\eta^{-})\!+\!I(\eta^{2},\eta^{-},\eta^{-})] \ , \label{I4momenta}
\end{eqnarray}
\begin{eqnarray}
I_{5}(p,k)&=&\int\!\frac{d^{4}pd^{4}k}{(2\pi)^8}\frac{1}{(p^{2}+\eta^{+})(p^{2}+\eta^{-})(k^{2}+\eta^{2})(k^{2}+\eta^{+})(k^{2}+\eta^{-})(q^{2}+\eta^{+})(q^{2}+\eta^{-})} \nonumber\\
&=&\frac{1}{(2b)^4}\left[-4I(\eta^{2},\eta^{+},\eta^{+})\!+\!8I(\eta^{2},\eta^{+},\eta^{-})\!-\!4I(\eta^{2},\eta^{-},\eta^{-})\!+\!2I(\eta^{+},\eta^{+},\eta^{+})\right. \nonumber\\
&& \ \ \ \ \ \ \ \ \ \left.-2I(\eta^{+},\eta^{+},\eta^{-})\!-\!2I(\eta^{+},\eta^{-},\eta^{-})\!+\!2I(\eta^{-},\eta^{-},\eta^{-})\right] \ . \label{I5momenta}
\end{eqnarray}

By taking the results (\ref{I4momenta}) and (\ref{I5momenta}) into (\ref{third diagram}), we get (\ref{diagram 3 integrals}).

For the fourth two-loop diagram, Eq. (\ref{diagram 4}),
\begin{eqnarray}
I_4&=&8g^{2}a^{2}b^{2}\int\!\frac{d^{4}pd^{4}k}{(2\pi)^8}E(p)E(k)B(q)+16g^{2}a^{2}b^{4}\int\!\frac{d^{4}pd^{4}k}{(2\pi)^8}E(p)B(k)B(q) \nonumber\\
&&+8g^{2}a^{2}b^{6}\int\!\frac{d^{4}pd^{4}k}{(2\pi)^8}B(p)B(k)B(q) \nonumber\\
&&-16g^{2}b^{2}\int\!\frac{d^{4}pd^{4}k}{(2\pi)^8}E(p)B(k)A(q)k^{2}q^{2}-8g^{2}b^{4}\int\!\frac{d^{4}pd^{4}k}{(2\pi)^8}B(p)B(k)A(q)q^{4} \nonumber\\
&&-16g^{2}m^{2}b^{2}\int\!\frac{d^{4}pd^{4}k}{(2\pi)^8}E(p)B(k)A(q)k^{2}-8g^{2}m^{2}b^{4}\int\!\frac{d^{4}pd^{4}k}{(2\pi)^8}B(p)B(k)A(q)q^{2} \nonumber\\
&=&8g^{2}a^{2}b^{2}I_{6}(p,k)+16g^{2}a^{2}b^{4}I_{7}(p,k)+8g^{2}a^{2}b^{6}I_{8}(p,k)-16g^{2}b^{2}I_{9}(p,k) \nonumber\\
&&-8g^{2}b^{4}I_{10}(p,k)-16g^{2}m^{2}b^{2}I_{11}(p,k)-8g^{2}m^{2}b^{4}I_{12}(p,k) \ , \label{fourth diagram}
\end{eqnarray}
with
\begin{eqnarray}
I_{6}(p,k)&=&\int\!\frac{d^{4}pd^{4}k}{(2\pi)^8}\frac{1}{(p^{2}+\eta^{2})(k^{2}+\eta^{2})(q^{2}+\eta^{2})(q^{2}+\eta^{+})(q^{2}+\eta^{-})} \nonumber\\
&=&\frac{1}{(2b)^2}[-4I(\eta^{2},\eta^{2},\eta^{2})\!+\!2I(\eta^{2},\eta^{2},\eta^{+})\!+\!2I(\eta^{2},\eta^{2},\eta^{-})] \ , \label{I6momenta}
\end{eqnarray}
\begin{eqnarray}
I_{7}(p,k)&=&\int\!\frac{d^{4}pd^{4}k}{(2\pi)^8}\frac{1}{(p^{2}+\eta^{2})(k^{2}+\eta^{2})(k^{2}+\eta^{+})(k^{2}+\eta^{-})(q^{2}+\eta^{2})(q^{2}+\eta^{+})(q^{2}+\eta^{-})} \nonumber\\
&=&\frac{1}{(2b)^4}\left[16I(\eta^{2},\eta^{2},\eta^{2})\!-\!16I(\eta^{2},\eta^{2},\eta^{+})\!-\!16I(\eta^{2},\eta^{2},\eta^{-})\!+\!4I(\eta^{2},\eta^{+},\eta^{+})\right. \nonumber\\
&& \ \ \ \ \ \ \ \ \ \left.+8I(\eta^{2},\eta^{+},\eta^{-})\!+\!4I(\eta^{2},\eta^{-},\eta^{-})\right] \ , \label{I7momenta}
\end{eqnarray}
\begin{eqnarray}
I_{8}(p,k)\!\!&=&\!\!\!\!\int\!\frac{d^{4}pd^{4}k}{(2\pi)^8}\frac{1}{(p^{2}\!+\!\eta^{2})\!(p^{2}\!+\!\eta^{+})\!(p^{2}\!+\!\eta^{-})\!(k^{2}\!+\!\eta^{2})\!(k^{2}\!+\!\eta^{+})\!(k^{2}\!+\!\eta^{-})\!(q^{2}\!+\!\eta^{2})\!(q^{2}\!+\!\eta^{+})\!(q^{2}\!+\!\eta^{-})} \nonumber\\
&=&\frac{1}{(2b)^6}\left[-64I(\eta^{2},\eta^{2},\eta^{2})\!+\!96I(\eta^{2},\eta^{2},\eta^{+})\!+\!96I(\eta^{2},\eta^{2},\eta^{-})\!-\!48I(\eta^{2},\eta^{+},\eta^{+})\right. \nonumber\\
&& \ \ \ \ \ \ \ \ \ \ \left.-96I(\eta^{2},\eta^{+},\eta^{-})\!-\!48I(\eta^{2},\eta^{-},\eta^{-})\!+\!8I(\eta^{+},\eta^{+},\eta^{+})\!+\!24I(\eta^{+},\eta^{+},\eta^{-})\right. \nonumber\\
&& \ \ \ \ \ \ \ \ \ \ \left.+24I(\eta^{+},\eta^{-},\eta^{-})\!+\!8I(\eta^{-},\eta^{-},\eta^{-})\right] \ , \label{I8momenta}
\end{eqnarray}
\begin{eqnarray}
I_{9}(p,k)&=&\int\!\frac{d^{4}pd^{4}k}{(2\pi)^8}\frac{k^{2}}{(p^{2}+\eta^{2})(k^{2}+\eta^{2})(k^{2}+\eta^{+})(k^{2}+\eta^{-})(q^{2}+\eta^{2})} \nonumber\\
&=&\int\!\frac{d^{4}pd^{4}k}{(2\pi)^8}\frac{1}{(p^{2}+\eta^{2})(k^{2}+\eta^{+})(k^{2}+\eta^{-})(q^{2}+\eta^{2})}-\eta^{2}I_{6}(p,k) \nonumber\\
&=&\frac{1}{(2b)^2}[4\eta^{2}I(\eta^{2},\eta^{2},\eta^{2})\!-\!2\eta^{+}I(\eta^{2},\eta^{2},\eta^{+})\!-\!2\eta^{-}I(\eta^{2},\eta^{2},\eta^{-})] \ , \label{I9momenta}
\end{eqnarray}
\begin{eqnarray}
I_{10}(p,k)&=&\int\!\frac{d^{4}pd^{4}k}{(2\pi)^8}\frac{q^{2}}{(p^{2}+\eta^{2})(p^{2}+\eta^{+})(p^{2}+\eta^{-})(k^{2}+\eta^{2})(k^{2}+\eta^{+})(k^{2}+\eta^{-})(q^{2}+\eta^{2})} \nonumber\\
&=&\int\!\frac{d^{4}pd^{4}k}{(2\pi)^8}\frac{1}{(p^{2}+\eta^{2})(p^{2}+\eta^{+})(p^{2}+\eta^{-})(k^{2}+\eta^{2})(k^{2}+\eta^{+})(k^{2}+\eta^{-})}-\eta^{2}I_{7}(p,k) \nonumber\\
&=&\frac{1}{(2b)^{4}}\left[16J(\eta^{2}\!,\!\eta^{2})\!-\!16J(\eta^{2}\!,\!\eta^{+})\!-\!16J(\eta^{2}\!,\!\eta^{-})\!+\!4J(\eta^{+}\!,\!\eta^{+})\!+\!8J(\eta^{+}\!,\!\eta^{-})\!+\!4J(\eta^{-}\!,\!\eta^{-})\right. \nonumber\\
&& \ \ \ \ \ \ \ \ \left.-16\eta^{2}I(\eta^{2},\eta^{2},\eta^{2})\!+\!16\eta^{2}I(\eta^{2},\eta^{2},\eta^{+})\!+\!16\eta^{2}I(\eta^{2},\eta^{2},\eta^{-})\!-\!4\eta^{2}I(\eta^{2},\eta^{+},\eta^{+})\right. \nonumber\\
&& \ \ \ \ \ \ \ \ \left.-8\eta^{2}I(\eta^{2},\eta^{+},\eta^{-})\!-\!4\eta^{2}I(\eta^{2},\eta^{-},\eta^{-})\right] \ , \label{I10momenta}
\end{eqnarray}
\begin{eqnarray}
I_{11}(p,k)&=&\int\!\frac{d^{4}pd^{4}k}{(2\pi)^8}\frac{k^{2}}{(p^{2}+\eta^{2})(k^{2}+\eta^{2})(k^{2}+\eta^{+})(k^{2}+\eta^{-})q^{2}(q^{2}+\eta^{2})} \nonumber\\
&=&\int\!\frac{d^{4}pd^{4}k}{(2\pi)^8}\frac{1}{(p^{2}+\eta^{2})(k^{2}+\eta^{+})(k^{2}+\eta^{-})q^{2}(q^{2}+\eta^{2})} \nonumber\\
&&-\eta^{2}\int\!\frac{d^{4}pd^{4}k}{(2\pi)^8}\frac{1}{(p^{2}+\eta^{2})(k^{2}+\eta^{2})(k^{2}+\eta^{+})(k^{2}+\eta^{-})q^{2}(q^{2}+\eta^{2})} \nonumber\\
&=&\frac{1}{(2b\eta)^2}\left[-4\eta^{2}I(\eta^{2},\eta^{2},\eta^{2})\!+\!2\eta^{+}I(\eta^{2},\eta^{2},\eta^{+})\!+\!2\eta^{-}I(\eta^{2},\eta^{2},\eta^{-})\!+\!4\eta^{2}I(\eta^{2},\eta^{2},0)\right. \nonumber\\
&& \ \ \ \ \ \ \ \ \ \ \left.-2\eta^{+}I(\eta^{2},\eta^{+},0)\!-\!2\eta^{-}I(\eta^{2},\eta^{-},0)\right] \ , \label{I11momenta}
\end{eqnarray}
\begin{eqnarray}
I_{12}(p,k)&=&\int\!\frac{d^{4}pd^{4}k}{(2\pi)^8}\frac{1}{(p^{2}+\eta^{2})(p^{2}+\eta^{+})(p^{2}+\eta^{-})(k^{2}+\eta^{2})(k^{2}+\eta^{+})(k^{2}+\eta^{-})(q^{2}+\eta^{2})} \nonumber\\
&=&I_{7}(p,k) \ . \label{I12momenta}
\end{eqnarray}

Plugging (\ref{I6momenta})-(\ref{I12momenta}) into (\ref{fourth diagram}) gives (\ref{diagram 4 integrals}).

Note that, unlike the previous integrals appearing in $I_1$, $I_2$ and $I_3$, $I_{6}(p,k)$ and $I_{9}(p,k)$ are log divergent, and even if we use (\ref{I(x,y,z)}), the final result for $I_4$ is not finite. 

In order to renormalize the divergent part of effective potential at two loops, we adopt the same strategy used in \cite{Jones,Espinosa,Martin}. As we are working with renormalized parameters, we just minimally subtract the subdivergence terms of the two-loop integrals, diagram by diagram, rather than compute separately a set of one-loop diagrams with counterterm insertions. Using this procedure, we do not need to calculate the renormalization constants necessary to cancel the $(1/\epsilon^2)$ and $(1/\epsilon)$ poles, and so the renormalized two-loop effective potential is written down by replacing the integrals calculated above by the $\epsilon$-independent part of the functions $\hat{I}(x,y,z)$ and $\hat{J}(x,y)$:
\begin{eqnarray}
\hat{J}(x,y)&=&J(x,y)+\frac{1}{\kappa\epsilon}(xJ(y)+yJ(x)) \ , \nonumber\\
\hat{I}(x,y,z)&=&I(x,y,z)-\frac{1}{\kappa\epsilon}(J(x)+J(y)+J(z)) \ . \label{IJ hat}
\end{eqnarray}

With these equations, and recalling (\ref{diagram 1 integrals}), (\ref{diagram 2 integrals}), (\ref{diagram 3 integrals}) and (\ref{diagram 4 integrals}), the two-loop effective potential is given by
\begin{eqnarray}
{\mathcal V}_{eff}^{(2)}&=&(I_{1}+h.c.)+(I_{2}+h.c.)+I_{3}+I_{4} \nonumber\\
&=&2g^{2}a^{2}\left[-4\hat{I}(\eta^{2},\eta^{2},\eta^{2})+3\hat{I}(\eta^{+},\eta^{+},\eta^{+})+\hat{I}(\eta^{+},\eta^{-},\eta^{-})\right] \nonumber\\
&&+\frac{8g^{2}a^{2}\eta^{-}}{\eta^2}\left[-\hat{I}(\eta^{2},\eta^{2},\eta^{+})+\hat{I}(\eta^{2},\eta^{2},\eta^{-})\right] \nonumber\\
&&+8g^{2}m^{2}\left[-2\hat{I}(\eta^{2},\eta^{2},0)+\frac{\eta^+}{\eta^2}\hat{I}(\eta^{2},\eta^{+},0)+\frac{\eta^-}{\eta^2}\hat{I}(\eta^{2},\eta^{-},0)\right] \nonumber\\
&&+2g^{2}\!\left[\!-\!4\hat{J}(\eta^{2}\!,\!\eta^{2})\!+\!4\hat{J}(\eta^{2}\!,\!\eta^{+})\!+\!4\hat{J}(\eta^{2}\!,\!\eta^{-})\!-\!\hat{J}(\eta^{+}\!,\!\eta^{+})\!-\!2\hat{J}(\eta^{+}\!,\!\eta^{-})\!-\!\hat{J}(\eta^{-}\!,\!\eta^{-})\right] \ . \label{V2 integrals}
\end{eqnarray}

This is the expression for the two-loop effective potential in terms of the integrals $\hat{I}$ and $\hat{J}$, given in (\ref{IJ hat}), and the finite part of this gives the renormalized two-loop effective potential (\ref{V2 final}).


\end{document}